\documentclass[12pt]{iopart}
\usepackage{iopams}  

\def\be{\begin{equation}}
\def\ee{\end{equation}}
\def\ba{\begin{eqnarray}}
\def\ea{\end{eqnarray}}
\def\bea{\begin{eqnarray}}
\def\eea{\end{eqnarray}}
\def\bean{\begin{eqnarray*}}
\def\eean{\end{eqnarray*}}
\def\bes{\numparts}
\def\ees{\endnumparts}
\def\btd{{\widetilde d}}

\newcommand{\A}{{\mathcal{A}}}

\newcommand{\tal}{{\widetilde \alpha}}

\newcommand{\td}{{\widetilde d}}

\newcommand{\MSbar}{\overline{\rm MS}}  

\newcommand{\tK}{{\widetilde K}}
\newcommand{\cM}{{\cal M}}
\usepackage{graphics}
\usepackage{graphicx}
\usepackage{dcolumn}
\usepackage{bm}
\usepackage{epsfig}
\usepackage{graphicx,color}
\usepackage{multirow}
\usepackage{xcolor}

\begin{document}

\title[]{Infrared-suppressed QCD coupling and the hadronic contribution to muon g-2}

\author{Gorazd Cveti\v{c}}
\address{Department of Physics, Universidad T{\'e}cnica Federico Santa Mar{\'\i}a, Casilla 110-V, Valpara{\'\i}so, Chile}

\author{Reinhart K\"ogerler}
\address{Department of Physics, Universit\"at Bielefeld, 33501 Bielefeld, Germany}
\vspace{10pt}
\begin{indented}
\item[]July 2020 
\end{indented}

\begin{abstract}
A variant of QCD with the coupling suppressed in the infrared (IR) regime, as suggested by large-volume lattice calculations of the Landau-gauge gluon and ghost dressing functions, is considered. The coupling is further restricted by the condition of approximate coincidence with perturbative QCD in the high momentum regime, and by the $\tau$-lepton semihadronic decay rate in the intermediate momentum regime, the rate which is evaluated by a renormalon-motivated resummation method. The obtained coupling turns out to be free of Landau singularities. The $D=4,6$ condensate values of the Adler function are then extracted by application of the Borel sum rules to the OPAL and ALEPH (V+A)-channel data of $\tau$-decay, and the corresponding V-channel condensate values are deduced as well. We then show that the correct value of the hadronic vacuum polarization contribution to the muon anomalous magnetic moment, $a_{\mu}^{\rm had(1)}$, is reproduced by regularizing the $D=4,6$ OPE terms in the V-channel Adler function with IR-regularization masses ${\cal M}_{D/2} \lesssim 1 \ {\rm GeV}$, suggesting the internal consistency of the presented QCD framework.
\end{abstract}

\vspace{2pc}

\maketitle

\section{Introduction}
\label{sec:intr}
Perturbative QCD (pQCD) cannot be applied to the studies of phenomenology in the intermediate ($|Q^2| \sim 1 \ {\rm GeV}^2$) and infrared ($|Q^2| < 1 \ {\rm GeV}^2$) regimes of the $Q^2$-complex plane,\footnote{We use the notation $Q^2 \equiv -q^2 = -(q^0)^2 + {\vec q}^2$, with $q$ being the 4-momentum in a considered physical process.} because the pQCD running coupling  $a(Q^2) \equiv \alpha_s(Q^2)/\pi$ has singularities within or close to such regimes. The problematic singularities appear in the spacelike IR-regime, $0 < Q^2 < \Lambda^2_{\rm Lan.}$ where $\Lambda^2_{\rm Lan.} \sim 0.1$-$1 \ {\rm GeV}^2$ is the branching scale of the Landau cut. These singularities lead, above all, to practical difficulties of evaluation of $a(Q^2)$ and of QCD processes at such $Q^2$. 

A way out of this problem consists in regaining a correct analytic behaviour by ``analytizing'' the running coupling, i.e., by replacing the pQCD coupling $a(Q^2)$ by another coupling \textcolor{black}{$\A(Q^2)$ which} has the aspired analyticity properties and could, at least in principle, be used for (quasi)perturbative evaluation of low-energy observables.

In Ref.~\cite{3dAQCD} we have constructed such an improved coupling and demonstrated its compatibility with intermediate-energy observables. Here we somewhat refine this construction and apply it in addition to quantities determined by even lower energies.

\section{Construction of $\A(Q^2)$}
\label{sec:consA}

Our coupling $\A(Q^2)$ is based on dispersive methods and determined mainly by two demands: I.) it should approach the pQCD coupling $a(Q^2)$ for $Q^2 \to \infty$; and II.) it should be compatible with lattice results at very low $Q^2$. Let us go into more details now.

For the behaviour of $\A(Q^2)$ in the IR regime we proceed as follows. We start from the general defining relation for the running pQCD coupling $a (Q^2) \equiv \alpha_s(Q^2)/\pi$:
\be
a (Q^2)  =  a (\Lambda^2) \frac{Z_{\rm gl}^{(\Lambda)}(Q^2) Z_{\rm gh}^{(\Lambda)}(Q^2)^2}{Z_1 ^{(\Lambda)}(Q^2)^2},
\label{alatt}
\ee  
where $Z_{\rm gl}$, $Z_{\rm gh}$ are the dressing functions of the gluon and ghost propagator, respectively, and $Z_1$ is the gluon-ghost-ghost vertex renormalization constant.
In the Landau gauge, in which large-volume lattice calculations are performed, $Z_1^{(\Lambda)}(Q^2)=1$ to all orders \cite{Taylor}. The resulting formula for the running coupling is particularly convenient for lattice calculations since single particle correlation functions (full propagators) in the Landau gauge are most easily accessible with that technique. In this way, a lattice coupling $\A_{\rm latt}(Q^2)$ can be defined
\be
\A_{\rm latt}(Q^2)  \equiv  \A_{\rm latt}(\Lambda^2) Z_{\rm gl}^{(\Lambda)}(Q^2) Z_{\rm gh}^{(\Lambda)}(Q^2)^2 \ ,
\label{Alatt}
\ee
where the $Z$'s result from large-volume lattice simulations (with lattice spacing $1/\Lambda$). $\A_{\rm latt}(Q^2)$ and our coupling $\A(Q^2)$ include both perturbative and nonperturbative contributions. We will require our coupling $\A(Q^2)$ [a low-energy extension of the pQCD coupling $a(Q^2)$] to agree qualitatively with $\A_{\rm latt}(Q^2)$ in the IR-regime 
\be
\A_{\rm latt}(Q^2) =  \A(Q^2) + \Delta \A_{\rm NP}(Q^2) \ ,
\label{AlattA}
\ee
\textcolor{black}{where $\Delta \A_{\rm NP}$  is regarded as a restricted (see below) nonperturbative difference between $\A_{\rm latt}$ and our $\A$.}
Recent large-volume lattice results \cite{LattcoupNf02,Lattcoupb,Lattcoupc} indicate that $\A_{\rm latt}(Q^2) \sim Q^2$ at $Q^2 \to 0$. We will assume that there is no finetuning at $Q^2 \to 0$; this leads to
\be 
\A(Q^2) \sim Q^2 \quad {\rm and} \quad \Delta \A_{\rm NP}(Q^2) \sim Q^2
\quad ({\rm at} \; Q^2 \to 0).
\label{noft}
\ee
A further result of lattice calculations, which we will use in the following, is that $\A_{\rm latt}(Q^2)$ (at real positive $Q^2$) shows a local maximum at $Q^2 \sim 0.1 \ {\rm GeV}^2$.

A note on renormalization schemes is in order here. The mentioned lattice calculations have been performed within the (lattice) MiniMOM (MM) scheme. Consequently, we also work within that scheme, but with the squared momenta rescaled to the usual $\MSbar$-like scaling: $Q^2=Q^2_{\rm latt} (\Lambda_{\MSbar}/\Lambda_{\rm MM})^2 \approx Q^2_{\rm latt}/1.9^2$.  We call this rescaled scheme the ``Lambert MM'' scheme (LMM). The name is motivated by the fact that, for the underlying perturbative coupling $a(Q^2)$, as well as for its  spectral function $\rho_a(\sigma) \equiv {\rm Im} \ a(-\sigma - i \epsilon)$, we use for calculational efficiency an explicit expression \cite{GCIK2011,3dAQCD} in terms of the Lambert function $W_{\pm 1}(z(Q^2))$. Here, $z(Q^2)=-(\Lambda_{L}/Q^2)^{\beta_0/c_1}/(c_1 e)$, and the coupling $a(Q^2)$ is in the LMM scheme which has the first two $\beta$ scheme coefficients $c_j=\beta_j/\beta_0$ equal to the known MM coefficients \cite{MiniMOM} (with $N_f=3$)
\be
c_2=9.2970 (4.4711), \; c_3=71.4538 (20.9902)\ ,
\label{c2c3}
\ee
where in parentheses the values in the $\MSbar$ scheme are given. 
The Lambert scale $\Lambda_{L}$ can be determined numerically from the value of $\alpha_s(M_Z^2;\MSbar)$. For example, when using the recent world average \textcolor{black}{values $\alpha_s(M_Z^2;N_f=5;\MSbar)=0.1179 \pm 0.0010$ \cite{PDG2019}, we get for the $N_f=3$ regime: $\Lambda_{L}=0.1120^{+0.0051}_{-0.0049}$ GeV} (we use the five-loop $\MSbar$ $\beta$-function \cite{5lMSbarbeta} and the corresponding four-loop quark threshold matching \cite{4lquarkthresh}). \textcolor{black}{In our specific example, we will use $\alpha_s(M_Z^2;N_f=5;\MSbar)=0.1177$ which gives   $\Lambda_{L}=0.1110$ GeV.}

Having clarified the renormalization scheme, the coupling $\A(Q^2)$ will be constructed by the dispersion relation
\be
\A(Q^2) = \frac{1}{\pi} \int_{\sigma=M^2_{\rm thr}}^{\infty} \frac{d \sigma \rho_{\A}(\sigma)}{(\sigma + Q^2)} ,
\label{Adisp}
\ee
where $\rho_{\A}(\sigma) \equiv {\rm Im} \; \A(Q^2=-\sigma - i \varepsilon)$, and $M^2_{\rm thr}$ is a threshold scale expected to be $\sim 0.1 \ {\rm GeV}^2$ [$\sim (2 m_{\pi})^2$].
In Eq.~(\ref{Adisp}) we have to specify the corresponding discontinuity function $\rho_{\A}(\sigma)$ for the whole energy range $\sigma \in [M^2_{\rm thr},\infty)$. We do this in two steps:

  In the UV-regime (large positive $\sigma=-Q^2$), we demand that $\rho_{\A}(\sigma)$ tend to the underlying pQCD spectral function $\rho_a(\sigma)$ as dictated by the asymptotic freedom
\be
\rho_{\A}(\sigma) = \rho_a(\sigma) \qquad {\rm for} \; \sigma > M_0^2,
\label{rhoAa} \ee
where $M_0^2$ ($ \sim 1$-$10 \ {\rm GeV}^2$) denotes the onset of the perturbative regime.

In the remaining (IR) region ($M^2_{\rm thr} < \sigma < M_0^2$) the spectral function $\rho_{\A}(\sigma)$ is a priori unknown, and we have to make a physically motivated ansatz. This interval contributes in the dispersion integral to the part which we call $\Delta \A_{\rm IR}(Q^2)$, and  we decide to parametrize the latter quantity by means of a quasidiagonal Pad\'e $[M-1/M](Q^2)$. This specific choice is motivated by the highly efficient convergence properties of these approximants when $M$ increases \cite{Peris}. On the other hand, we need to keep the number of free adjustable parameters limited in order to avoid numerical instabilities during the adjustments. We take $M=3$
\be
\Delta \A_{\rm IR}(Q^2) = \frac{\sum_{n=1}^{2} A_n Q^{2n}}{\sum_{n=1}^3 B_n Q^{2n}}
=  \sum_{j=1}^{3} \frac{{\cal F}_j}{Q^2 + M_j^2}.
\label{PFM1M}
\ee
The second expression on the right-hand side is obtained by the partial-fraction decomposition of the Pad\'e, with free adjustable parameters ${\cal F}_j$ and $M_j$ ($j=1,2,3$). Together with Eq.~(\ref{rhoAa}) this implies
\be
\A(Q^2)=   \sum_{j=1}^3 \frac{{\cal F}_j}{(Q^2 + M_j^2)} + \frac{1}{\pi} \int_{M_0^2}^{\infty} d \sigma \frac{ \rho_a (\sigma) }{(Q^2 + \sigma)},
\label{AQ2} \ee
and the corresponding spectral function is
\be
\rho_{\A}(\sigma) =  \pi \sum_{j=1}^{3} {\cal F}_j \; \delta(\sigma - M_j^2)  + \Theta(\sigma - M_0^2) \rho_a (\sigma),
\label{rhoA}
\ee

Finally, we have to fix the seven as yet unspecified parameters [$M_j,{\cal F}_j$ ($j=1,2,3$) and $M_0^2$], and therefore we need seven appropriate conditions:

I.) Four conditions stem from the requirement that the coupling $\A$ at high momenta ($|Q^2|>\Lambda_L^2$) practically (i.e., up to high power corrections) coincides with the (underlying) pQCD coupling $a$: $\A(Q^2) - a(Q^2) \sim (\Lambda_{L}^2/Q^2)^N$, where $N$ is sufficiently high. We take $N=5$ which gives four conditions (cf.~Ref.~\cite{3dAQCD} for more details).

II.) The fifth condition is implied by the limiting behaviour $\A(Q^2) \sim Q^2$ for $Q^2 \to 0$, cf.~Eq.~(\ref{noft}).

III.) The sixth condition comes from the fact that for positive $Q^2$ the lattice coupling $\A_{\rm latt}(Q^2)$ has a maximum at $Q^2_{\rm max} \approx 0.135 \ {\rm GeV}^2$ [in the mentioned rescaled ``Lambert'' MM (LMM) scheme]; we require that our $\A(Q^2)$ achieves maximum at the same $Q^2_{\rm max} \approx 0.135 \ {\rm GeV}^2$.\footnote{We note that the last two conditions (II. and III.) are the only information that we take from lattice calculations.}

IV.) The final, seventh, condition is connected with the requirement that the use of the coupling $\A(Q^2)$ in QCD (we call this the $\A$QCD framework) should work well in the intermediate energy regime ($|Q^2| \sim 1 \ {\rm GeV}^2$). Specifically, it should reproduce the correct value of the canonical hadronic $\tau$-decay branching ratio $r^{(D=0)}_{\tau} \approx 0.20$ \cite{ALEPH2}. This is the QCD-part of the hadronic $\tau$-decay ratio into nonstrange hadrons, with all higher-twist ($D \not=0$) and nonzero quark mass contributions subtracted. In Appendix, a summarized analysis and evaluation of this quantity within $\A$QCD is given, where it is evaluated with the renormalon-motivated model of Ref.~\cite{renmod}. The equality of the theoretical value of  $r^{(D=0)}_{\tau}$ with the experimentally preferred value $r^{(D=0)}_{\tau} \approx 0.20$ leads to the seventh condition.

\begin{figure}[htb] 
\begin{minipage}[b]{.49\linewidth}
  \centering\includegraphics[width=80mm]{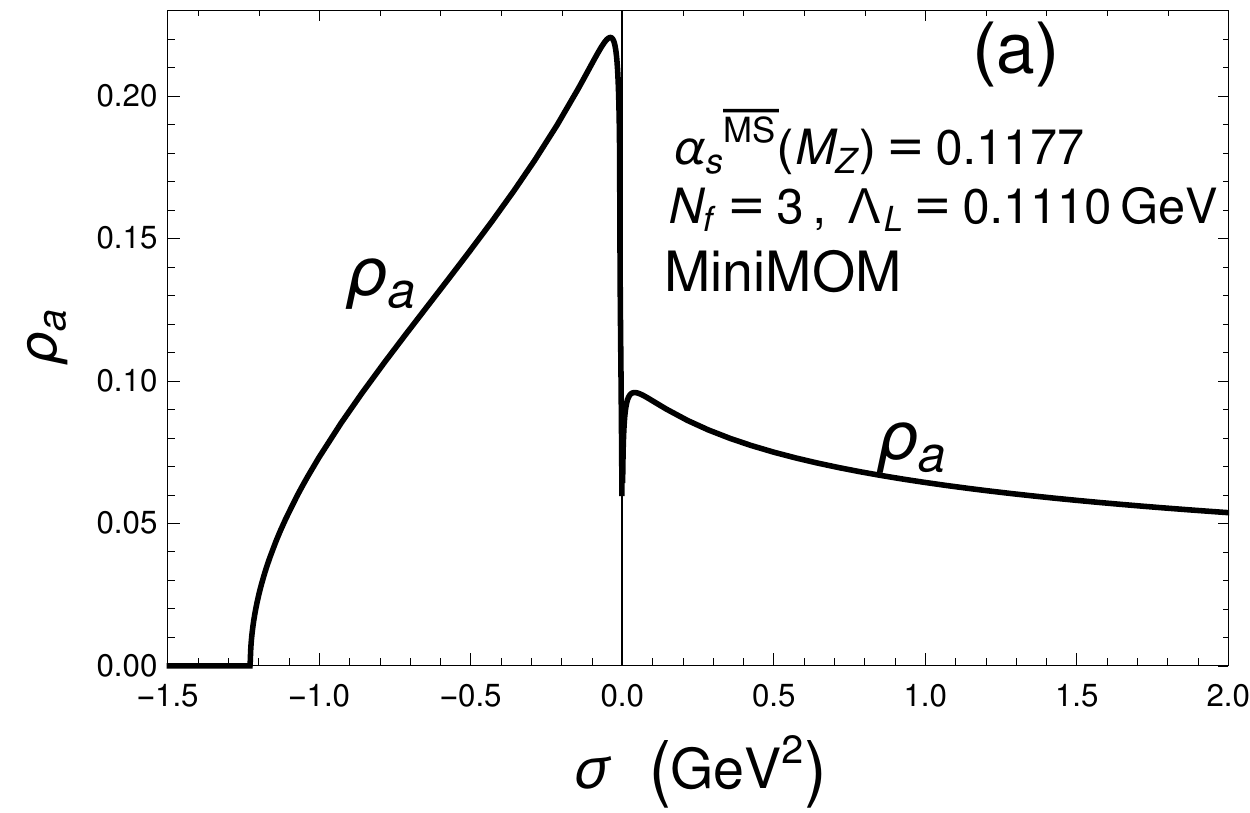}
  \end{minipage}
\begin{minipage}[b]{.49\linewidth}
  \centering\includegraphics[width=80mm]{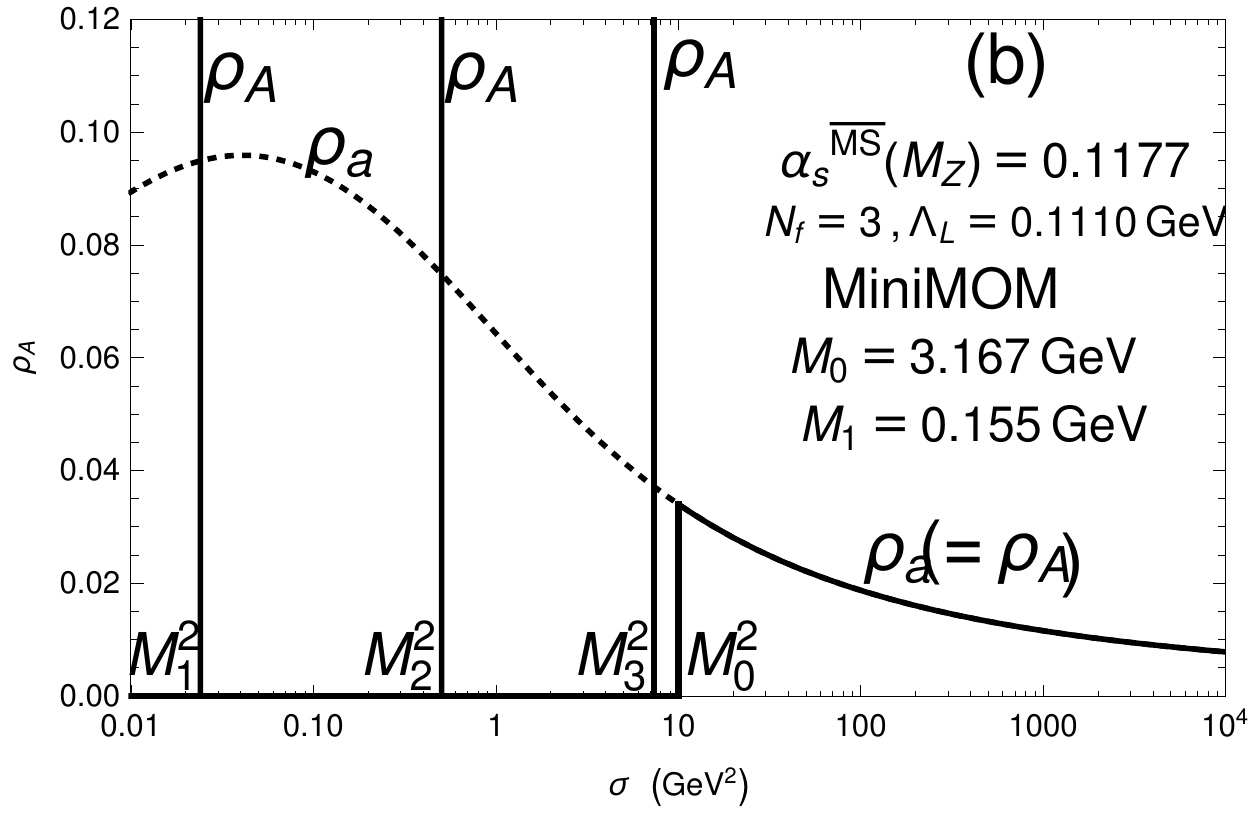}
\end{minipage}
\vspace{-0.4cm}
\caption{(a) The spectral function $\rho_a (\sigma) = {\rm Im} \; a (Q^2=-\sigma - i \epsilon)$ in the 4-loop LMM scheme, $\sigma$ is on linear scale; (b) $\rho_{\A}(\sigma) =  {\rm Im} \; \A(Q^2=-\sigma - i \epsilon)$, where $\sigma > 0$ is on logarithmic scale. The delta function at $M_1^2$ is negative (shown as positive for convenience).}
\label{Figrho}
\end{figure}

\begin{figure}[htb] 
\centering\includegraphics[width=95mm]{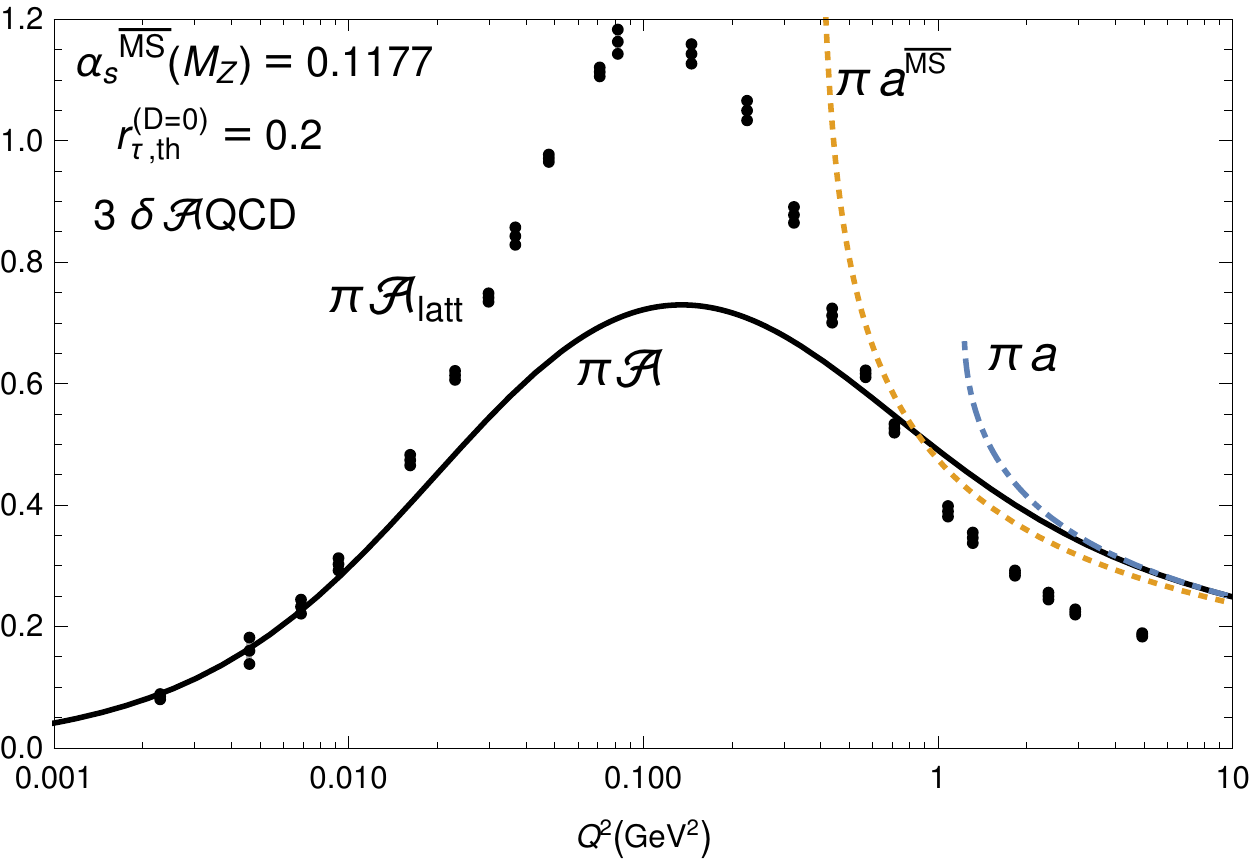}
\vspace{-0.4cm}
\caption{The considered $N_f=3$ holomorphic coupling $\pi \A$ (solid curve), the underlying LMM pQCD coupling $\pi a$ (dot-dashed curve), $\MSbar$ pQCD coupling ${\overline a}$ (dotted curve), for $Q^2>0$. Included are the large-volume lattice results $\pi \A_{\rm latt}$ \cite{LattcoupNf02} (points with bars), for which the momenta $Q^2$ were rescaled from the lattice MM to the LMM scheme: $Q^2=Q^2_{\rm latt} (\Lambda_{\MSbar}/\Lambda_{\rm MM})^2 \approx Q^2_{\rm latt}/1.9^2$. At large $Q^2 > 1 \ {\rm GeV}^2$, the (large-volume) lattice results are unreliable.}
\label{FigAa}
 \end{figure}

The seven conditions taken together lead us to obtain numerical values of the seven parameters of the coupling $\A(Q^2)$ Eq.~(\ref{AQ2}). \textcolor{black}{When we choose $\alpha_s(M_Z^2;\MSbar)=0.1177$ and  $r^{(D=0)}_{\tau, {\rm th}}=0.200$, we obtain \cite{GCRKext}:}
\begingroup \color{black}
\bes
\bea
M_0^2 &= & 10.033  \ {\rm GeV}^2 \; (M_0 \approx 3.167 \ {\rm GeV}); 
\label{M0}
\\
M_1^2 &=& 0.0240 \ {\rm GeV}^2 \; (M_1 \approx 0.155 \ {\rm GeV}), \quad {\cal F}_1  =  -0.00813 \ {\rm GeV}^2,
\label{M1}
\\
M_2^2&=&0.506  \ {\rm GeV}^2 \; (M_2 \approx 0.712 \ {\rm GeV}), \quad {\cal F}_2 = 0.1313 \ {\rm GeV}^2,
\label{M2}
\\
M_3^2 &=& 7.358  \ {\rm GeV}^2 \; (M_3 \approx 2.713 \ {\rm GeV}), \quad {\cal F}_3 = 0.0740 \ {\rm GeV}^2.
\label{M3}
\eea
\ees  \endgroup
We see that all ${M}^2_j > 0$, therefore the resulting coupling $\A(Q^2)$ is holomorphic (i.e., without the Landau singularities) not by imposition, but as a result of the seven mentioned (physically-motivated) conditions. \textcolor{black}{In Fig.~\ref{Figrho} we present the underlying pQCD spectral function $\rho_a$ and the resulting spectral function $\rho_{\A}$. In Fig.~\ref{FigAa} the resulting coupling $\pi \A(Q^2)$ at low positive $Q^2$ is given. At $Q^2 \to 0$ the coupling behaves as $A(Q^2) = k Q^2$ with $k \approx 13.6 \ {\rm GeV}^{-2}$. The coupling agrees qualitatively with the lattice results, while the height of the peak depends significantly on the chosen reference value $\alpha_s(M_Z^2; \MSbar)$ both in our approach and in the lattice calculation.}

Somewhat different (but similar) results were obtained by us earlier \cite{3dAQCD}, where for the Adler function (see Sec.~\ref{sec:BSR} and Appendix) we took the nonresummed form: truncated series in $\A$QCD based on the first four terms of the pQCD expansion (\ref{dpt}).\footnote{\textcolor{black}{The Dirac delta function at $\sigma=M_3^2$ has an effect of simulating (parametrizing) a nonabrupt fall of $\rho_{\A}(\sigma)$ when $\sigma$ decreases below $M_0^2$.}} 
As a consequence, the evaluation of integrals in Eq.~(\ref{DAres}) [in contrast to $d(Q^2)_{D=0; res} $ of Eq.~(\ref{Dares})] is unambiguous for all spacelike $Q^2 \in \mathbb{C} \backslash (-\infty, -M_1^2]$, because no Landau singularities are encountered along the integration lines, \textcolor{black}{and the resulting Adler function  $d(Q^2)_{D=0; \A res}$ is a holomorphic function in the entire complex $Q^2$-plane with the exception of the negative semiaxis.}

\textcolor{black}{The values of parameters in Eqs.~(\ref{M0})-(\ref{M3}) change appreciably when the values of the input parameters $\alpha_s(M_Z^2; \MSbar)$ and $r_{\tau, {\rm th}}^{(D=0)}$ change. For example, when  $\alpha_s(M_Z^2; \MSbar)$ is increased  to $0.1181$, the extracted parameters are: $M_0 \approx 2.864$ GeV, $M_1=0.252$ GeV, $M_2=0.454$ GeV, $M_3= 2.442$ GeV; ${\cal F}_1=-0.0582 \ {\rm GeV}^2$; ${\cal F}_2=0.1716 \ {\rm GeV}^2$; ${\cal F}_3=0.0665 \ {\rm GeV}^2$.}

\section{Applications: I.~Borel sum rules for semihadronic $\tau$ decay}
\label{sec:BSR}

An important physical quantity, essential for the analysis of several QCD processes (e.g., hadronic $\tau$-decays, cf.~Appendix) is the Adler function ${\cal D}(Q^2)$ defined by
\be  {\cal D}(Q^2) \equiv  - 2 \pi^2 \frac{d \Pi(Q^2)}{d \ln Q^2},
\label{Ddef} \ee
where $\Pi(Q^2)$ is the general vacuum polarization function, i.e., current correlation function. The OPE expansion of the (full V+A channel) Adler function is
\be
{\cal D}_{\rm V+A}(Q^2)
=  1 + d(Q^2)_{D=0} + 2 \pi^2 \sum_{n \geq 2}
 \frac{ n  \langle O_{2n} \rangle_{\rm V+A}}{(Q^2)^n}.
\label{DVA}
\ee
The leading-twist ($D=0$) contribution $d(Q^2)_{D=0}$, sometimes also called Adler function, and its evaluation in $\A$QCD with the renormalon-motivated model of Ref.~\cite{renmod}, are explained in Appendix. The nonperturbative higher-twist terms ($D \geq 4$) include the corresponding vacuum condensates.

The Adler function can be used in general sum rules. Namely, by choosing any holomorphic function $g(Q^2)$ one can derive from it the sum rule
  \be
\int_0^{\sigma_{\rm max}} d \sigma g(-\sigma) \omega_{\rm exp}(\sigma)  =
-i \pi  \oint d Q^2 g(Q^2) \Pi_{\rm th}(Q^2)  \ ,
\label{sr1}
\ee
where the integration on the right-hand side is performed along the circle $|Q^2|=\sigma_{\rm max}$  (with $\sigma_{\rm max} \leq m^2_{\tau}$); $\omega(\sigma)$ is the spectral function of $\Pi(Q^2)$ along the cut,
$\omega(\sigma) \equiv 2 \pi \; {\rm Im} \ \Pi(Q^2=-\sigma - i \epsilon)$, which is measured. Integration by parts on the right-hand side of Eq.~(\ref{sr1}) leads to a form which involves the Adler function ${\cal D}(Q^2)$.

\begin{figure}[htb] 
\begin{minipage}[b]{.49\linewidth}
  \centering\includegraphics[width=80mm]{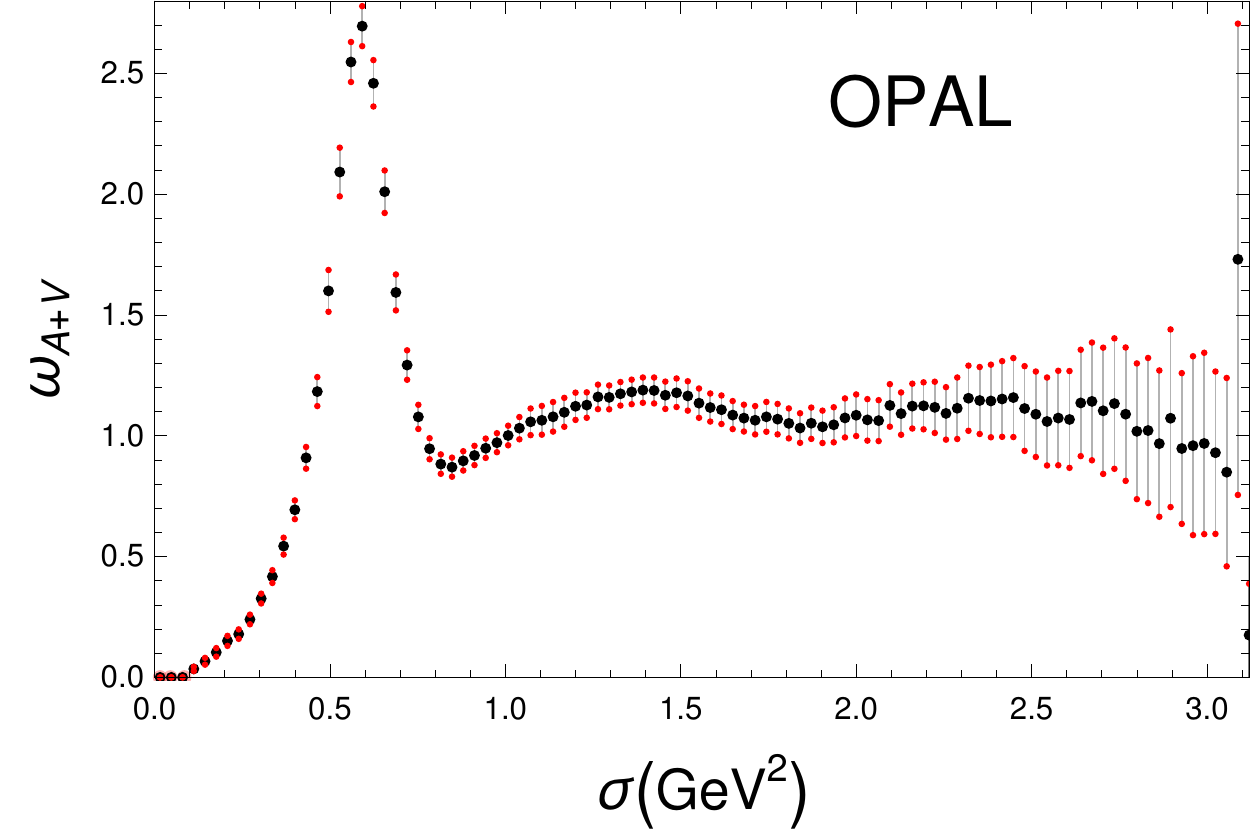}
  \end{minipage}
\begin{minipage}[b]{.49\linewidth}
  \centering\includegraphics[width=80mm]{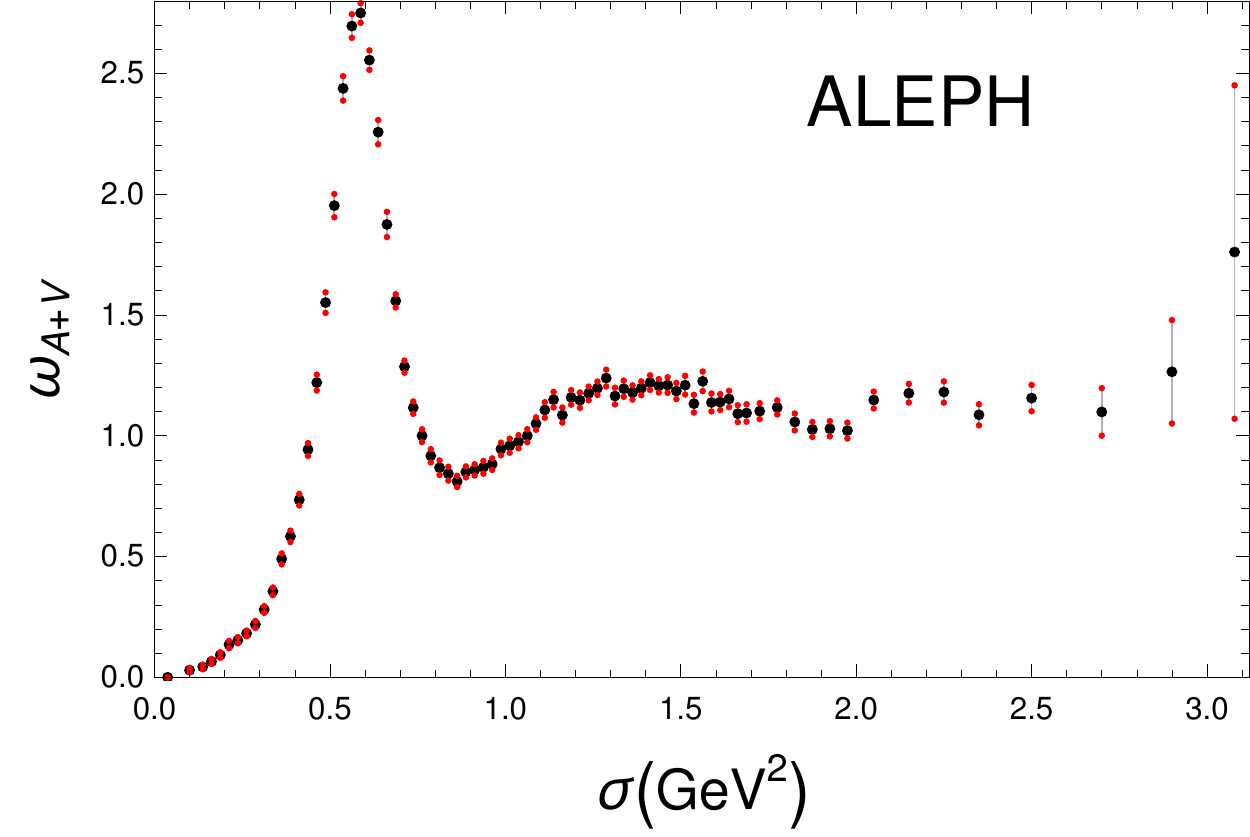}
\end{minipage}
\vspace{-0.4cm}
\caption{(a) The spectral function $\omega_{{\rm V+A}}(\sigma)$, measured by OPAL (left-hand) and by ALEPH Collaboration (right-hand), without the pion peak contribution. We take $\sigma_{\rm max}=3.136$ and $2.80 \ {\rm GeV}^2$ for OPAL and ALEPH, respectively.}
\label{FigOmega}
\end{figure}

\begin{figure}[htb] 
\begin{minipage}[b]{.49\linewidth}
  \centering\includegraphics[width=78mm]{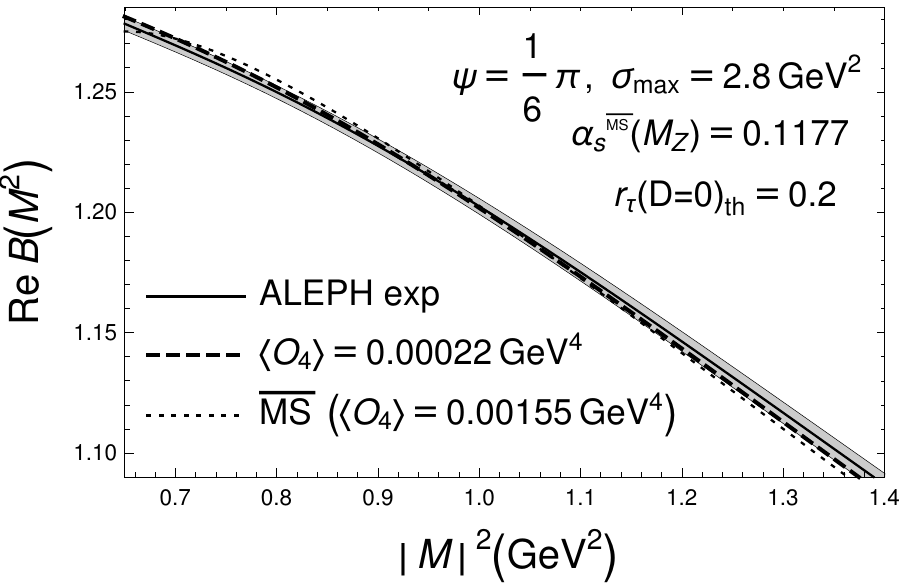}
  \end{minipage}
\begin{minipage}[b]{.49\linewidth}
  \centering\includegraphics[width=78mm]{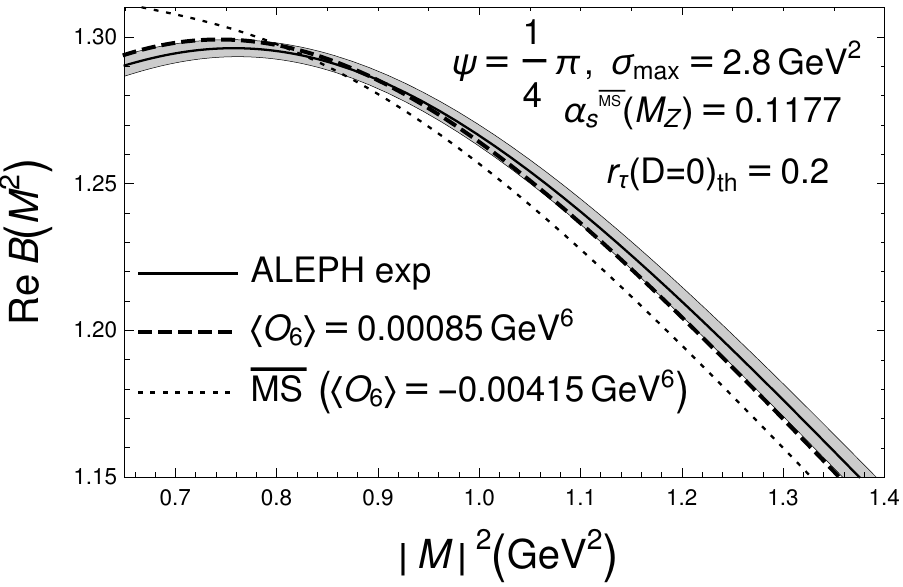}
\end{minipage}
\vspace{-0.4cm}
\caption{Borel transforms ${\rm Re} B(M^2)$ along the rays $M^2 = |M^2| \exp(i \Psi)$ with $\Psi=\pi/6$ (left-hand) and $\Psi = \pi/4$ (right-hand), as a function of $|M^2|$, fitted to the ALEPH data. \textcolor{black}{The grey band represents exprimental results [left-hand side of Eq.~(\ref{sr1})], the solid line is the middle of this band.}}
\label{FigPi64}
\end{figure}

\begin{figure}[htb] 
  \centering\includegraphics[width=95mm]{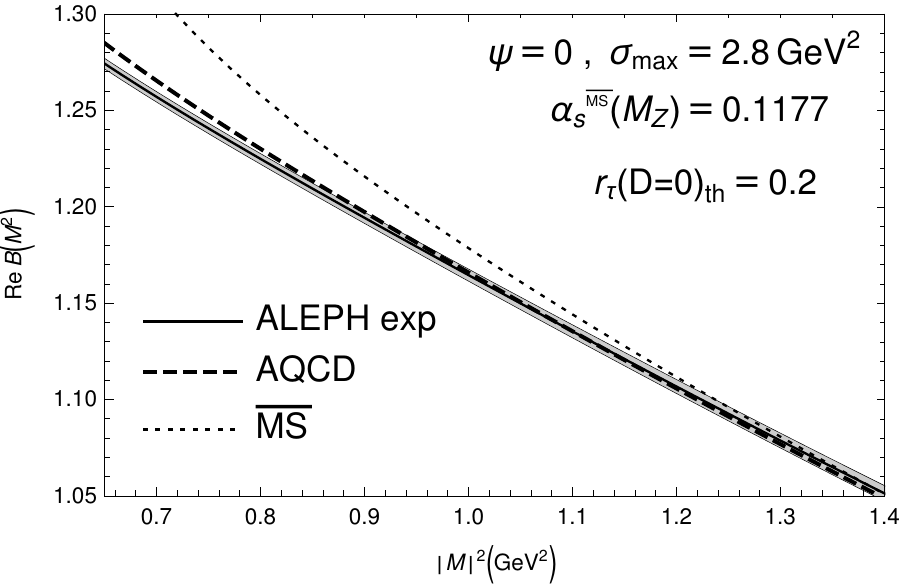}
\vspace{-0.4cm} 
 \caption{Analogous to the previous Figures, but now the Borel transforms $B(M^2)$ are for real $M^2 > 0$.}
\label{FigPsi0}
 \end{figure}

The specific case of Borel (or: Laplace) sum rules is obtained if one chooses $g(Q^2) = \exp(Q^2/M^2)/M^2$, where $M^2$ denotes a complex parameter (Borel scale), and in Eq.~(\ref{sr1}) only the real parts are considered \cite{Ioffe}. The corresponding integrals on the right-hand side are usually denoted as $B_{\rm (th)}(M^2)$. If in the sum in Eq.~(\ref{DVA}) we take only two terms ($n=2,3$), then for $M^2=|M^2| \exp(i \pi/6)$ and $M^2=|M^2| \exp(i \pi/4)$ the Borel sum rules allow us to extract the condensate values $\langle O_4 \rangle_{\rm V+A}$ and $\langle O_6 \rangle_{\rm V+A}$, respectively, from the measured $\tau$-decay spectral function $\omega_{\rm exp}(\sigma) = 2 \pi {\rm Im} \Pi_{{\rm V+A}}(-\sigma - i \epsilon)$ as obtained from OPAL \cite{OPAL} and ALEPH \cite{ALEPH2} experiments, cf.~Fig.~\ref{FigOmega}.
In Figs.~\ref{FigPi64} we present the results of the fit to the ALEPH data, for Borel transforms with $M^2 = |M^2| \exp(i \Psi)$ with $\Psi=\pi/6$ and $\pi/4$. We can see that the adjustment of the values of $\langle O_4 \rangle$ and $\langle O_6 \rangle$, respectively, gives a good fit to the central experimental curve when the described $\A$QCD evaluation is used in the $D=0$ part of the Adler function (\ref{DVA}) in the LMM scheme [see Eq.~(\ref{DAres})]. When applying the  $\MSbar$ pQCD approach, the $D=0$ part of Adler function is calculated in the $\MSbar$ scheme (at complex $Q^2$) according to Eq.~(\ref{Dares}) with the characteristic functions for $\MSbar$ from Ref.~\cite{renmod}. We can see in Figs.~\ref{FigPi64} that the $\MSbar$ pQCD approach gives worse fit. Finally, in Fig.~\ref{FigPsi0} the curves for the theoretical Borel transforms for real positive $M^2$ (i.e., $\Psi=0$) are given, with the condensate values of $\langle O_4 \rangle$ and $\langle O_6 \rangle$ obtained from the aforementioned fits, and compared with the experimental ALEPH values. The $\A$QCD prediction is significantly better than the $\MSbar$ pQCD.

The gluon condensate is directly related to $\langle O_4 \rangle_{{\rm V+A}}$: $\langle a GG \rangle = 6 \langle O_4 \rangle_{{\rm V+A}} + 0.00199 \ {\rm GeV}^4$.

Combining the fits for OPAL and ALEPH data yields, using the described $\A$ for fixed \textcolor{black}{$\alpha_s(M_Z^2;\MSbar)=0.1177$ and $r_{\tau}^{(D=0)}=0.200$} (for comparison we include the results with $\MSbar$ pQCD coupling)
\begingroup \color{black}
\bes
\bea
\langle O_4 \rangle_{{\rm V+A}} & = & (+0.00028 \pm 0.00016) \ {\rm GeV}^4
\label{O4}
\\ 
\Rightarrow  \; \langle a G G \rangle &=&  (+0.00364 \pm 0.00097) \ {\rm GeV}^4 ,
\label{aGG}
\\
\langle O_6 \rangle_{{\rm V+A}} & = & (+0.00074 \pm 0.00021) \ {\rm GeV}^6.
\label{O6}
\eea
\bea
\langle O_4 \rangle_{{\rm V+A},\MSbar} & = & (+0.00173 \pm 0.00024) \ {\rm GeV}^4, 
\label{O4MS}
\\
\langle O_6 \rangle_{{\rm V+A},\MSbar} & = & (-0.00451 \pm 0.00040) \ {\rm GeV}^6. 
\label{O6MS}
\eea
\ees \endgroup
Cross-check of consistency can be performed, comparing the theoretical (predetermined by the coupling $\A$, cf.~Appendix) and the experimental values of $r^{(D=0,  \sigma_{\rm max})}_{\tau}$:
\bes
\bea
\lefteqn{r^{(D=0, \sigma_{\rm max})}_{\tau,{\rm exp}}=}
\nonumber\\
&=& 2 \int_{0}^{\sigma_{\rm max}} \frac{d \sigma}{\sigma_{\rm max}} \left( 1 - \frac{\sigma}{\sigma_{\rm max}} \right)^2 \left( 1 + 2  \frac{\sigma}{\sigma_{\rm max}} \right) \omega_{\rm exp}(\sigma)-1 + 12 \pi^2 \frac{\langle O_6 \rangle_{{\rm V+A}}}{\sigma^3_{\rm max}}
\nonumber\\
& =& \begingroup \color{black} 0.201 \pm 0.006 \ \quad \rm (OPAL) \; vs \; 0.201 \; \rm (th.) \endgroup
\label{rtauexpa}
\\
& =& \begingroup \color{black} 0.211 \pm 0.003 \ \quad \rm (ALEPH) \; vs \; 0.213 \; \rm (th.) \endgroup
  \label{rtauexpb}
\eea
\ees 
We see that there is a consistency of the $\A$QCD results. We have $\sigma_{\rm max}=3.136 \ {\rm GeV}^2$ and $2.80 \ {\rm GeV}^2$, and \textcolor{black}{$\langle O_6 \rangle_{\rm V+A}=+0.00085$ and $+0.00063 \ {\rm GeV}^6$, for OPAL and ALEPH, respectively. We recall that $r^{(D=0, m_{\tau}^2)}_{\tau,{\rm th}} =0.200$.}

\section{Applications: II.~V-channel Adler function and muon $g-2$}
\label{sec:DV}

We can now perform a further consistency check of our $\A$QCD [called $3\delta$ $\A$QCD to address the specific ansatz Eq.~(\ref{rhoA})], by applying it to a quantity which is determined by QCD at even lower energies, namely the anomalous magnetic moment of muon, or more specifically, to the leading order hadronic vacuum polarization (had(1)) contribution to this moment, $(g_{\mu}/2-1)^{\rm had(1)} \equiv a_{\mu}^{\rm had(1)}$. This quantity is experimentally deducible to a high accuracy from the precise measurements of the cross section $e^+ e^- \to \gamma^{\ast} \to$ hadrons, the recent values are \textcolor{black}{\cite{Davier:2019can}}\footnote{\textcolor{black}{Recently, the BMW-Collaboration \cite{Borsanyi:2020mff} obtained from lattice calculation significantly higher values $10^{10} \times a_{\mu; {\rm exp}}^{\rm had(1)} = 712.4 \pm 4.5$, indicating that no new physics beyond SM (beyond QCD) is required to explain the directly measured value of full $a_{\mu}$ \cite{PDG2019}. On the other hand, another lattice calculation \cite{Lehner:2020crt} indicates that the lattice results have higher statistical uncertainties than assumed in \cite{Borsanyi:2020mff}, which would avoid tension with the result Eq.~(\ref{amuhad1}) based on the measurements of  $e^+ e^- \to \gamma^{\ast} \to$ hadrons.}}
\be
10^{10}   \times a_{\mu; {\rm exp}}^{\rm had(1)} \approx \color{black} 694 \pm 4 \ .
\label{amuhad1}
\ee 
For its theoretical evaluation, one needs the correlation function of the V-channel currents. Therefore, first the evaluation of the (full) V-channel Adler function 
\bea
{\cal D}_{\rm V}(Q^2) &\equiv&  - 4 \pi^2 \frac{d \Pi_{\rm V}(Q^2)}{d \ln Q^2} 
= d(Q^2)_{D=0} + {\cal D}(Q^2)^{\rm (NP)}
\nonumber\\
 &=&  d(Q^2)_{D=0} + 1 + 2 \pi^2 \sum_{n \geq 2}
 \frac{ n 2 \langle O_{2n} \rangle_{\rm V}}{(Q^2)^n}
\label{DV}
\eea
is needed. The leading-twist contribution $d(Q^2)_{D=0}$ is the same as in the previously considered (V+A)-channel case (cf.~ Appendix). The $D=4$ condensate values here are known from the (V+A)-channel because
$\langle O_4 \rangle_{\rm V} =  \langle O_4 \rangle_A  = (1/2) \langle O_4 \rangle_{{\rm V+A}}$.
On the other hand, for $D=6$, a sum rule analysis of the (V-A)-channel \cite{GPRS} gives (the average from OPAL and ALEPH data)
\be
\langle O_6 \rangle_{{\rm V-A}} = (-0.00465 \pm 0.00126) \ {\rm GeV}^6,
\label{O6VmA}
\ee
Therefore, taking into account that $\langle O_6 \rangle_{\rm V} = (1/2) ( \langle O_6 \rangle_{{\rm V+A}} +  \langle O_6 \rangle_{{\rm V-A}})$, we obtain in our case of $3 \delta$ $\A$QCD [with \textcolor{black}{$\alpha_s(M_Z^2;\MSbar)=0.1177$ and $r_{\tau}^{(D=0)}=0.200$}]
\bes
\bea
\langle O_4 \rangle_{\rm V} &=& \textcolor{black}{ (+0.00014 \pm 0.00008)} \ {\rm GeV}^4,
\label{O4Vres}
\\
\langle O_6 \rangle_{\rm V} &=& \textcolor{black}{ (-0.00196 \pm 0.00064)} \ {\rm GeV}^6,  
\label{O6Vres} 
\eea \ees
The nonperturbative (NP) part of V-channel Adler function (\ref{DV}), since applied here to very low $Q^2$-values, clearly has to be ``IR-regularized.''
We do this by two regularization masses $\cM_2$ and $\cM_3$ in the following way:
\bes
\bea
\lefteqn{
  {\cal D}_{\rm V}(Q^2)^{\rm (NP)}  =  1 + 2 \pi^2 \sum_{n \geq 2}
\frac{ n 2 \langle O_{2n} \rangle_{\rm V}}{(Q^2)^n}
\label{NPOPE}
}
\\
& = &
1 + 4 \pi^2
\left[ \frac{2 \langle O_4 \rangle_{\rm V}}{ (Q^2+ \cM_2^2)^2 } 
            + \frac{\left( 3 \langle O_6 \rangle_{\rm V} + 4 \cM_2^2 \langle O_4 \rangle_{\rm V} \right)}{\textcolor{black}{(Q^2 + \cM_3^2)^3}} \right]
\label{NPOPEreg}
\eea \ees
We note that similarly regularized higher-twist expressions were used in the analyses of Bjorken Sum Rule in \cite{KTG}.
The IR-regularization masses $\cM_2$ and $\cM_3$ are expected to be real positive and $\lesssim 1$ GeV, reflecting the scales of the NP regime in QCD. The expression (\ref{NPOPEreg}) is written in such a way that in the limit of large $|Q^2|$ it gives the correct first two terms of the sum in Eq.~(\ref{NPOPE}). At $Q^2 \to 0$ we have, on the other hand
  \be
  {\cal D}_{\rm V}(0) = 0 \; \Rightarrow  {\cal D}_{\rm V}(0)^{\rm (NP)} =0.
  \label{deepIR1}
  \ee
  The above implication is valid because the (resummed) $D=0$ part $d(Q^2)_{D=0}$ in $3 \delta$ $\A$QCD, Eq.~(\ref{DAres}) goes to zero when $Q^2 \to 0$ because $\A(Q^2) \sim Q^2 \to 0$. The condition ${\cal D}_{\rm V}(0)^{\rm (NP)} =0$ implies
\be
\cM_3^2  = \left[ \frac{(-3) \langle O_6 \rangle_{\rm V} - 4 \cM_2^2 \langle O_4 \rangle_{\rm V}}{\frac{1}{4 \pi^2} + 2   \langle O_4 \rangle_{\rm V}/\cM_2^4} \right]^{1/3}.
  \label{M32}
\ee 

Now we turn to the theoretical evaluation of $a_{\mu}^{\rm had(1)}$. It is given by
\bea
a_{\mu}^{\rm had(1)} &=& \frac{\alpha_{em}^2}{3 \pi^2} \int_0^{\infty} \frac{ds}{s} K(s) R_{\gamma,{\rm data}}(s),
\label{ahada}
\\
\!\!\!\!\!\!\!\!\!\!\!\!\!\!\!\!\!\!\!\!\!\!\!\!\!\!\!\!
{\rm with:} \qquad \;\;
K(s) &=&  \int_0^1 dx \frac{x^2 (1-x)}{x^2 + \frac{s}{m_{\mu}^2} (1-x)},
\label{Ks}
\eea
and $R_{\gamma,{\rm data}}(s) = 4 \pi k_f \; {\rm Im} \Pi_{\rm V}(-s-i \epsilon)$. Further, $k_f = 3 \sum_{f} Q_f^2$  ($k_f=2$ for $N_f=3$).

Using Cauchy theorem, $a_{\mu}^{\rm had(1)}$ Eq.~(\ref{ahada})  can be expressed in terms of the V-channel Adler function (\ref{DV})
\be
a_{\mu}^{\rm had(1)} =  \frac{\alpha_{em}^2}{3 \pi^2} \! \int_0^{1} \! \frac{dx}{x} (1-x) (2-x) {\cal D}_{\rm V} \left( Q^2\!=\!m^2_{\mu} \frac{x^2}{(1-x)} \right).
\label{ahadb}
\ee
We use now: (I) our $3 \delta$ $\A$QCD for the (renormalon-motivated) evaluation of the $D=0$ contribution to ${\cal D}_{\rm V}( Q^2)$ as given by Eq.~(\ref{DAres}); (II) the resummed OPE expression (\ref{NPOPEreg}) for the $D \geq 4$ (NP) contribution to  ${\cal D}_{\rm V}( Q^2)$; (III) and the obtained values of $\langle O_4 \rangle_{\rm V}$ and $\langle O_6 \rangle_{\rm V}$ Eqs.~(\ref{O4Vres})-(\ref{O6Vres}).  When evaluating the expression (\ref{ahadb}) in this way and requiring that the result reproduces the experimental value (\ref{amuhad1}), we can numerically extract the allowed values of the regularization masses $\cM_2$ and $\cM_3$. We obtain (the uncertainties from various sources are separated)
\begingroup \color{black}
\bes
\bea
\cM_2 &= & \left[ 0.384^{+0.019}_{-0.040} (\delta \langle O_4 \rangle_{\rm V})^{-0.019}_{+0.014} (\delta \langle O_6 \rangle_{\rm V}) \right] \ {\rm GeV},
\label{cM2} 
\\
\cM_3 & = & \left[ 0.730^{-0.012}_{+0.016} (\delta \langle O_4 \rangle_{\rm V})^{-0.055}_{+0.042} (\delta \langle O_6 \rangle_{\rm V}) \right] \ {\rm GeV}.
\label{cM3} 
\eea \ees \endgroup
This means that both IR-regularization masses are in the physically expected range of values, which we consider as a further consistency check of our approach. \textcolor{black}{If only the $D=0$ part $d(Q^2)_{D=0}$ were used for ${\cal D}_{\rm V}(Q^2)$, the integration Eq.~(\ref{ahadb}) would give $10^{10} \times a_{\mu, D=0}^{\rm had(1)} \approx 423$, i.e., about $61 \%$ of the required value Eq.~(\ref{amuhad1}). The extracted central values of Eqs.~(\ref{cM2})-(\ref{cM3}) change only very little when $10^{10} \times \delta a_{\mu; {\rm exp}}^{\rm had(1)}$ varies by $\pm 4$ according to Eq.~(\ref{amuhad1}): $\delta \cM_2 = \pm 0.7$ MeV and $\delta  \cM_3= \pm 0.3$ MeV. When the higher value $10^{10} \times a_{\mu; {\rm exp}}^{\rm had(1)} = 712.4$ is used (the central value of the prediction of Ref.~\cite{Borsanyi:2020mff}), the extracted values incease only slightly: $\delta \cM_2 = 3.1$ MeV and $\delta \cM_3=1.3$ MeV.}

\textcolor{black}{We have also performed our analysis for various other input values of  $\alpha_s(M_Z^2;\MSbar)$ and  $r_{{\tau}, {\rm th}}^{(D=0)}$  ($\equiv r_{{\tau}, {\rm th}}^{(D=0, m_{\tau}^2)}$), in order to investigate the stability of the results. Details of the results will be presented in the extended version of this work, Ref.~\cite{GCRKext}.}

\section{Conclusions}
\label{sec:concl}

A QCD coupling $\A(Q^2)$ was constructed by dispersive methods, in the lattice MiniMOM scheme (rescaled to the usual $\Lambda_{\MSbar}$ scale convention). Mathematica programs for the evaluation of the coupling is available online \cite{MathPrg}. This coupling defines a version of ($\A$)QCD which has several attractive features:

\noindent
(a) At high momenta $|Q^2| > 1 \ {\rm GeV}^2$, the coupling $\A(Q^2)$ reproduces the pQCD results, because there it practically coincides with the underlying pQCD coupling $a(Q^2) \equiv \alpha_s(Q^2)/\pi$.

\noindent
(b) At very low momenta $|Q^2| \lesssim 0.1 \ {\rm GeV}^2$, $\A(Q^2)$ goes to zero as $\sim Q^2$, as suggested by high-volume lattice results.

\noindent
(c) At intermediate momenta $|Q^2| \sim 1 \ {\rm GeV}^2$, $\A(Q^2)$ reproduces the well measured physics of semihadronic $\tau$-lepton decay.

\noindent
(d) As a byproduct of the construction, $\A(Q^2)$ possesses an attractive holomorphic behaviour in the complex $Q^2$-plane, the behaviour qualitatively shared by QCD spacelike physical quantities.

\noindent
(e) Several successful applications of $\A(Q^2)$ were made in low-$|Q^2|$ phenomenology, including the correct reproduction of the value of muon $(g-2)_{\rm exp}^{\rm had(1)}$.

Other holomorphic couplings have been introduced and applied in QCD phenomenology by various authors, among them \cite{ShS,MS,Sh1Sh2,BMS,APTappl1,APTappl2,Nest2,2dAQCD,PTBMF,Boucaud,Luna,Pelaez,Siringo,Mirjalili}. Further, spacelike QCD observables can be evaluated also by applying dispersive methods directly to them \cite{MSS1,MSS2,DeRafael,MagrGl,MagrTau,mes2,Nest3a,Nest3b}.

Further details will be presented in the extended version of this work, Ref.~\cite{GCRKext}.


\vspace{0.4cm}

\noindent
{\it Acknowledgments:\/} The work of G.C. was supported in part by the Fondecyt (Chile) Grant No.~1180344.

\appendix

\section{Nonstrange hadronic $\tau$-decay ratio}
\label{app:taudec}

We present here the expressions for the branching ratio $r_{\tau}^{(D=0)}$ of the $\tau$-decay into nonstrange hadrons, first in pQCD and then in $\A$QCD. In $r_{\tau}^{(D=0)}$, the contributions of nonzero quark masses and higher-twist terms ($D>0$) are subtracted, and it is normalized in the canonical way: $r_{\tau,{\rm pt}}^{(D=0)} = a + {\cal O}(a^2)$. This quantity can be expressed theoretically in terms of the Adler function $d(Q^2)_{D=0}$ (see also Sec.~\ref{sec:BSR})
\be
r^{(D=0)}_{\tau, {\rm th}} = \frac{1}{2 \pi} \int_{-\pi}^{+ \pi}
d \phi \ (1 + e^{i \phi})^3 (1 - e^{i \phi}) \
d(Q^2=m_{\tau}^2 e^{i \phi})_{D=0},
\label{rtaucont}
\ee
where $d(Q^2)_{D=0} = -1 - 2 \pi^2 d \Pi(Q^2)_{D=0}/d \ln Q^2$ is the massless Adler function. If we replace in Eq.~(\ref{rtaucont}) $m_{\tau}^2$ by $\sigma_{\rm max}$ ($\leq m_{\tau}^2$), we obtain the quantity $r^{(D=0, \sigma_{\rm max})}_{\tau, {\rm th}}$. The perturbation expansion of $r^{(D=0)}_{\tau, {\rm th}}$ is known up to ${\mathcal O}(a^4)$. Since $a(m^2_{\tau})$ is rather large ($\approx 0.312/\pi, 0.342/\pi$ in $\MSbar$ and LMM, respectively), one would wish to improve this approximation. Recently, by applying a renormalon-motivated model \cite{renmod}, we succeeded to extend the perturbation expansion formally to all orders in $a$
\be
d(Q^2)_{D=0; pt} =  a (Q^2) + \sum_{n=1}^{\infty} d_n a (Q^2)^{n+1},
\label{dpt}
\ee
It can be shown that this can be formally resummed with the corresponding characteristic functions  $G_D^{(j)}(t)$ \cite{renmod}
\bea
d(Q^2)_{D=0; res} &=& \int_0^1 \frac{dt}{t} \; G_D^{(-)}(t) a(t e^{-{\tK}} Q^2)
+ \int_1^{\infty} \frac{dt}{t} \; G_D^{(+)}(t) a(t e^{-\tK} Q^2)
\nonumber\\
&& + \int_0^1 \frac{dt}{t} \; G_D^{\rm (SL)}(t) \left[ a(t e^{-\tK} Q^2) - a(e^{-\tK} Q^2) \right],
\label{Dares}
\eea
where the characteristic functions are
\bea
G_D^{(-)}(t) &=&   \pi t^2 \left[{\td}_{2,1}^{\rm IR} - {\td}_{3,2}^{\rm IR} t \ln t  \right], \quad
G_D^{(+)}(t) =  \frac{\pi}{t} { \td}_{1,2}^{\rm UV} \ln t,
\label{GDpm}
\\
G_D^{\rm (SL)}(t) &=&  - { \tal} { \td}_{2,1}^{\rm IR} \frac{\pi t^2}{\ln t},
\label{GDSL} \eea 
and ${ \td}_{2,1}^{\rm IR}=-1.831$,  ${ \td}_{3,2}^{\rm IR}=11.05$, ${ \td}_{1,2}^{\rm UV}=0.005885$, ${ \tal}=-0.14$; ${ \tK}=-0.7704$: these are the renormalon-motivated parameters appearing in the Borel transform ${\rm B}[\btd](u)$ of an ``Adler''-related auxiliary quantity $\btd(Q^2; \mu^2)$.

It turns out that the correct resummation in $\A$QCD (``$\A$res'') is obtained from the pQCD resummation Eq.~(\ref{Dares}) by simply replacing $a \mapsto \A$ in all the integrands:
\bea
d(Q^2)_{D=0;\A res} &=& \int_0^1 \frac{dt}{t} \; G_D^{(-)}(t) \A(t e^{-{\tK}} Q^2)
+ \int_1^{\infty} \frac{dt}{t} \; G_D^{(+)}(t) \A(t e^{-\tK} Q^2)
\nonumber\\
&& + \int_0^1 \frac{dt}{t} \; G_D^{\rm (SL)}(t) \left[ \A(t e^{-\tK} Q^2) - \A(e^{-\tK} Q^2) \right],
\label{DAres}
\eea

By inserting $d(Q^2)_{D=0; \A res}$ into Eq.~(\ref{rtaucont}) and setting $r^{(D=0)}_{\tau, {\rm th}}=0.200$, we obtain the seventh condition for $\A(Q^2)$ and for the respective parameters.
    

\end{document}